\documentclass[twocolumn,floatfix,prl,showpacs,superscriptaddress]{revtex4}
\usepackage{graphicx}
\usepackage{amsbsy,amssymb,amsmath,bm}
\usepackage{graphicx,color,epsfig,rotate}

\begin{document}
\title{Calorimetric Evidence for a Fulde--Ferrell--Larkin--%
Ovchinnikov Superconducting State in the
Layered Organic Superconductor $\kappa$-(BEDT-TTF)$_2$Cu(NCS)$_2$}
\author{R. Lortz}
\affiliation{Department of Condensed Matter Physics, University of Geneva,
CH-1211 Geneva 4, Switzerland}
\author{Y. Wang}
\affiliation{Grenoble High Magnetic Field Laboratory, CNRS,
38042 Grenoble Cedex 9, France}
\author{A. Demuer}
\affiliation{Grenoble High Magnetic Field Laboratory, CNRS,
38042 Grenoble Cedex 9, France}
\author {P.H.M.~B\"ottger}
\affiliation {Hochfeld-Magnetlabor Dresden (HLD), Forschungszentrum
Dresden-Rossendorf, D-01314 Dresden, Germany}
\author {B.~Bergk}
\affiliation {Hochfeld-Magnetlabor Dresden (HLD), Forschungszentrum
Dresden-Rossendorf, D-01314 Dresden, Germany}
\author{G.~Zwicknagl}
\affiliation {Institut f\"ur Mathematische Physik, Technische
Universit\"at Braunschweig, D-38106 Braunschweig, Germany}
\author{Y.~Nakazawa}
\affiliation{Department of Chemistry, Osaka University, 1-1,
Machikaneyama, Toyonaka, Osaka, Japan}
\author{J. Wosnitza}
\affiliation {Hochfeld-Magnetlabor Dresden (HLD), Forschungszentrum
Dresden-Rossendorf, D-01314 Dresden, Germany}

\date{\today}

\begin{abstract}
The specific heat of the layered organic superconductor $\kappa$-%
(BEDT-TTF)$_2$Cu(NCS)$_2$, where BEDT-TTF is bisethylenedithio-%
tetrathiafulvalene, has been studied in magnetic fields
up to 28~T applied perpendicular and parallel to the superconducting
layers. In parallel fields above 21 T, the superconducting transition
becomes first order, which signals that the Pauli-limiting field is
reached. Instead of saturating at this field value, the upper critical
field increases sharply and a second first-order transition line
appears within the superconducting phase. Our results give strong
evidence that the phase, which separates the homogeneous superconducting
state from the normal state is a realization of a
Fulde-Ferrell-Larkin-Ovchinnikov state.

\end{abstract}
\pacs{74.70.Kn, 75.40.Cx, 74.25.Bt, 65.40.Ba}

\maketitle

The influence of high magnetic fields on the superconducting state is
of high technological and fundamental relevance. Applied aspects such
as high superconducting current densities as well as basic questions
on the nature of the superconducting state are of special interest.
In type-II spin-singlet
superconductors, a magnetic field destroys superconductivity in
two distinct ways: i.e., by orbital and Pauli-paramagnetic
pair-breaking effects. In general, both effects limit the maximum
upper critical field, $H_{c2}$, above which the normal-conducting state
is restored. In most type-II superconductors the orbital effect plays
the dominant role. If, however, under certain circumstances, the orbital
pair-breaking field, $H_{orb}$, is clearly larger than the Pauli
paramagnetic limit, $H_P$, the superconductor may enter a special,
spatially modulated, superconducting state at high magnetic field and
low temperatures. As predicted independently by Fulde and Ferrell
\cite{Fulde} and Larkin and Ovchinnikov \cite{Larkin} in 1964, in this
so-called FFLO (or LOFF) state superconductivity can survive even
above the Pauli limit by ``sacrificing'' parts of the material
volume to the normal state.

In the FFLO state, the Zeeman-split Fermi surfaces allow Cooper
pairing only with a finite center-of-mass momentum $\bm{q}$ resulting
in an oscillating part of the order parameter in real space with
wavelength of the order of the coherence length, $\xi$. For the FFLO
state to occur the so-called Maki parameter \cite{Mak64}, $\alpha =
\sqrt{2}H_{orb}/H_P$, should be larger than 1.8 \cite{Gru66} and the
superconductor needs to be in the clean limit with a mean-free path,
$\ell$, much larger than $\xi$. Not many superconductors fulfil these
conditions and early searches failed to observe the FFLO state.
In the 1990s, some reports claiming evidence for the FFLO state appeared
that later had to be revised. For more details see \cite{Bia03} and the
recent reviews \cite{Cas04,Mat07}.

Only recently, solid thermodynamic evidence for the existence of the
FFLO state has been put forward for the heavy-fermion compound CeCoIn$_5$
\cite{Bia02,Rad03}. Besides heavy-fermion superconductors the quasi-%
two-dimensional (2D) organic superconductors have been suggested as good
candidates for exhibiting the FFLO state \cite{Cas04,Mat07,Shi94,Shi97}.
For these the orbital pair breaking can be
greatly suppressed when applying the magnetic field parallel to the
highly conducting layers \cite{Shi94,Shi97,Bur94}. Indeed some signs,
but no thermodynamic proof, for an FFLO state in 2D organic superconductors
have been reported \cite{Sin00,Tan02,Uji06}. For $\lambda$-%
(BETS)$_2$GaCl$_4$ a kink in the thermal conductivity \cite{Tan02} and
for $\lambda$-(BETS)$_2$FeCl$_4$ dip structures in the resistance
\cite{Uji06} suggested the existence of a FFLO state (BETS is
bis\-ethylene\-dithio-tetraselenafulvalene).
For $\kappa$-(BEDT-TTF)$_2$Cu(NCS)$_2$ the existence of the FFLO phase
had been inferred from measurements sensitive to a loss in vortex
stiffness (BEDT-TTF is bis\-ethylene\-dithio-tetrathiafulvalene)
\cite{Sin00}. The observed features are, however, rather broad anomalies
that do not fit with our specific-heat results discussed below
and are, therefore, most probably not related to the FFLO state.

In this Letter, we present clear thermodynamic evidence that for
$\kappa$-(BEDT-TTF)$_2$Cu(NCS)$_2$ a narrow intermediate superconducting
state, most probably an FFLO state, evolves at high magnetic fields
applied parallel to the layers. At this Pauli-limited field region the
slope of the upper-critical-field line increases sharply and a first-order
transition appears below $H_{c2}$.

Single crystals of $\kappa$-(BEDT-TTF)$_2$Cu(NCS)$_2$ were grown by
the standard electrochemical-oxidation method as described elsewhere
\cite{Ura88}. The specific heat was measured by use of a miniaturized
relaxation technique \cite{Buq02,Heu06} in Geneva up to 14~T and in
the Grenoble High Magnetic Field Laboratory up to 28~T. In both cases,
the same calorimeter was used without removing the sample between the
experiments. The chip resistance and the thermal conductance of the
leads have been carefully calibrated up to 28~T using a capacitance
thermometer. Each relaxation provides about 1000 data points over a
temperature interval of 30-40\% above the base temperature, which has
been varied between 1.3 and 12~K. Data can be recorded during heating
and cooling, which allows to resolve hysteresis effects close to
first-order transitions. The merging of the upward and downward
relaxation data provides a highly reliable check of the accuracy of
this method.

The specific heat is a purely thermodynamic bulk quantity. Effects
related to flux pinning or vortex instabilities can thus be excluded
\cite{Lor06}. We precisely aligned
the sample for parallel-field orientation by slightly turning the cryostat
in the bore of the resistive coil while maximizing the superconducting
transition temperature in an intermediate field of 8~T. We found that
a good indicator for a perfect orientation was the absence of the
first-order vortex-melting transition in fields between 3 and 14~T
which is observable in slightly tilted fields \cite{Lor07}.

In order to determine the particular band-structure parameters of the
investigated specific-heat sample we measured the de Haas--van Alphen
(dHvA) effect by use of the cantilever-torque technique in a $^3$He
cryostat.

\begin{figure}
\centering
\includegraphics[width=0.44\textwidth,clip=true]{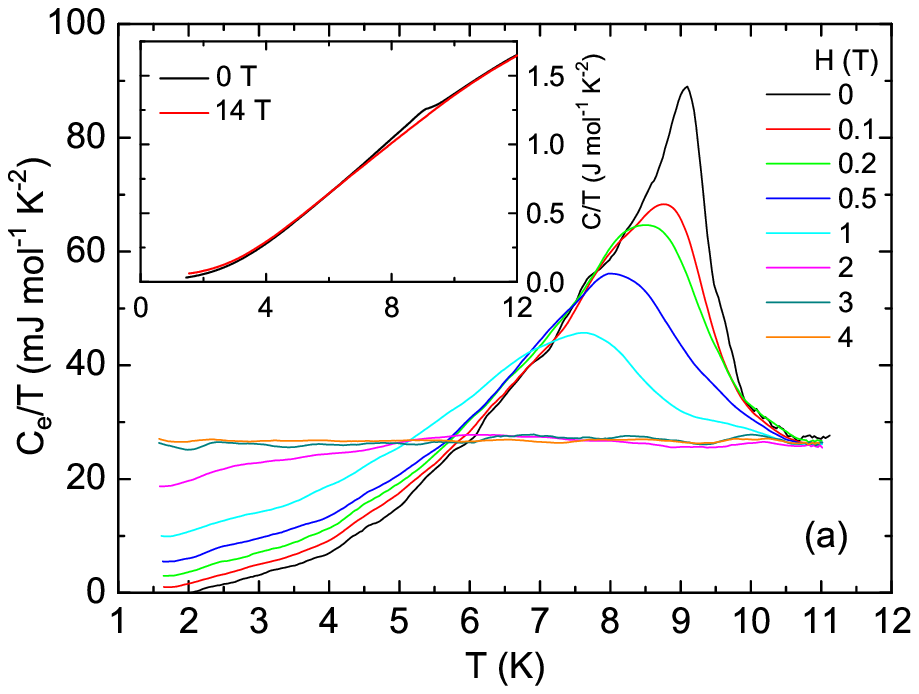}
\includegraphics[width=0.44\textwidth,clip=true]{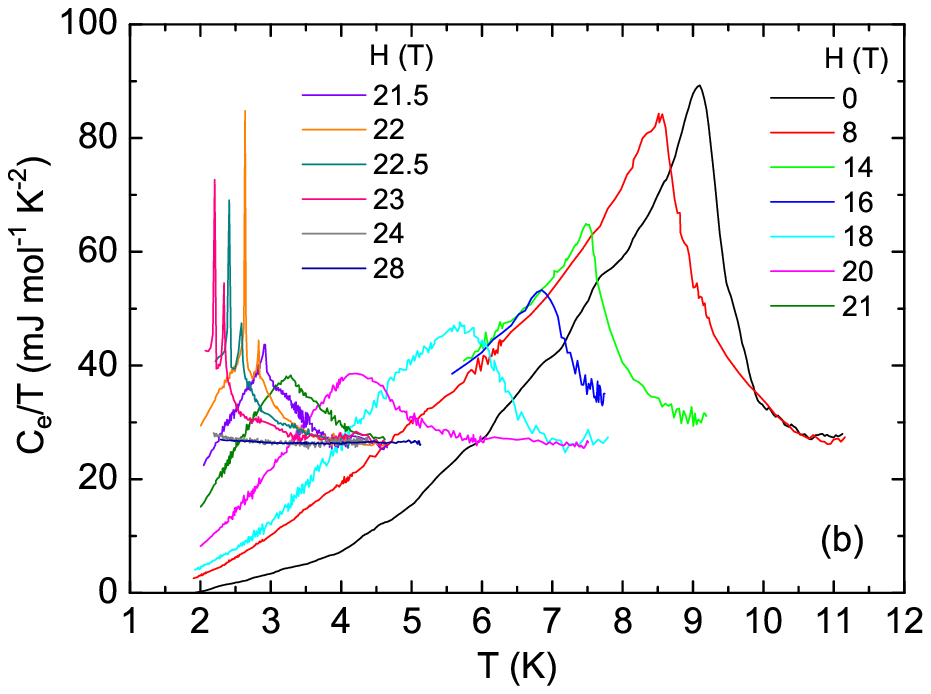}
\caption[]{(color) Temperature dependence of the electronic contribution
to the specific heat, $C_e/T$, of $\kappa$-(BEDT-TTF)$_2$Cu(NCS)$_2$
in magnetic fields applied perpendicular (a) parallel (b) to the
superconducting layers. Data taken in 14~T applied perpendicular
to the layers (inset) were used to separate the phonon contribution.}
\label{fig1ab}
\end{figure}

The inset of Fig.\ \ref{fig1ab}(a) shows specific-heat data taken in
0 and 14~T applied perpendicular to the superconducting layers. The
specific-heat anomaly, which represents only 5\% of the total specific
heat, is visible in zero field at $T_c = 9.1$~K. The curves merge for
all fields between 4 and 14~T which indicates that the upper critical
field is reached at about 4~T for this field orientation. The 14-T
data thus represent the normal state and allow us to extract the
Sommerfeld constant, $\gamma$, and to separate the phonon background.
We obtain $\gamma = 26(2)$~mJmol$^{-1}$K$^{-2}$ and a Debye temperature
of about 200~K. These values as well as the overall specific heat are
in very good agreement with earlier results \cite{And89,Mul02,Wos03}.
Subtraction of the phonon contribution, i.e., the 14-Tesla data minus
$\gamma T$, results in the electronic specific heat, $C_e$, shown for
fields applied perpendicular and parallel to the layers in Fig.\
\ref{fig1ab}(a) and \ref{fig1ab}(b), respectively. From the size of the
specific-heat jump $\Delta C=0.58$~Jmol$^{-1}$K$^{-2}$ we obtain the
ratio $\Delta C/(\gamma T_c)= 2.5(2)$, which, in accordance with
earlier reports \cite{Mul02,Wos03}, is larger than the BCS value
(1.43) and thus proves strong coupling.

The specific heats look rather different for the two field
orientations. For perpendicular fields the superconducting
transition is strongly broadened already in small fields due to
fluctuations. This can be explained by a field-induced finite-size
effect in layered superconductors \cite{Lor03}. Contrary to standard
superconductors in which $T_c$ is lowered in a field but essentially
remains sharp, the onset of the transition in $\kappa$-(BEDT-TTF)%
$_2$Cu(NCS)$_2$ is hardly influenced. Instead, each sign of
superconductivity continuously fades away while reaching $H_{c2}$.
A possible scenario may therefore be that in
$\kappa$-(BEDT-TTF)$_2$Cu(NCS)$_2$ only the phase coherence is lost
in magnetic fields, while Cooper pairs still exist above $H_{c2}$.

For fields applied parallel to the layers we find as well a broadening
of the superconducting transition, but this effect is less pronounced
and a clear shift of the specific-heat anomaly occurs. This resembles
more a standard behavior. The effect of fluctuations is
nevertheless also visible for this field orientation.

\begin{figure}
\centering
\includegraphics[width=0.44\textwidth,clip=true]{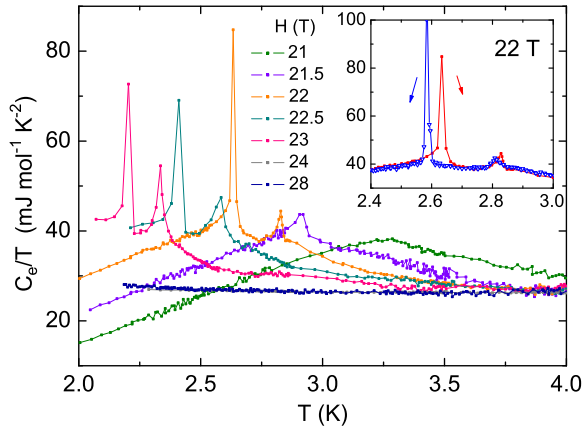}
\caption[]{(color) Details of $C_e/T$ of $\kappa$-(BEDT-TTF)$_2$Cu(NCS)$_2$
in high magnetic fields applied parallel to the superconducting layers.}
\label{figure2}
\end{figure}

Above $\sim$14~T, $T_c$ shifts rather rapidly with increasing field and
the transition becomes strongly broadened by fluctuations. At about
21~T, the anomaly in $C$ sharpens and at 21.5~T a spike in $C$ due to
the latent heat of a first-order transition appears (Fig.\ \ref{figure2}).
For higher fields up to 23~T, two sharp anomalies in $C$ clearly
prove the existence of an additional thermodynamic phase within the
superconducting state. The lower transition shows a well-%
resolvable temperature hysteresis ($\sim$0.05~K at 22~T), while the
main superconducting transition develops only a small hysteresis
$< 0.02$~K (inset of Fig.\ \ref{figure2}). We were able to follow the
two transitions up to 23~T, above which they left the temperature window
of our experiment.

\begin{figure}
\centering
\includegraphics[width=0.44\textwidth,clip=true]{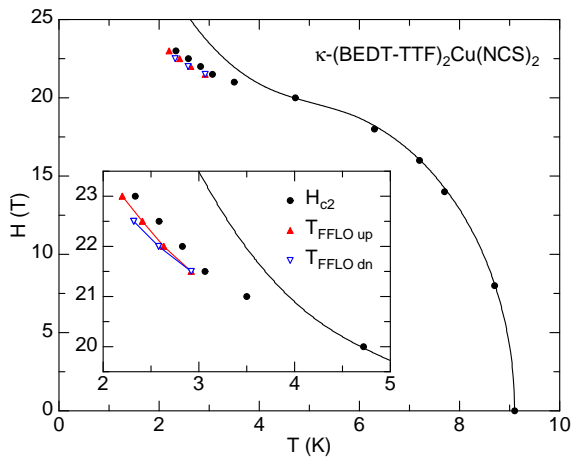}
\caption[]{(color) Phase diagram of $\kappa$-(BEDT-TTF)$_2$Cu(NCS)$_2$
for fields applied parallel to the superconducting layers.
The solid line represents the calculated $H_{c2}$ using the known
band-structure parameters. See the text for details.}
\label{figure3}
\end{figure}

The extracted magnetic phase diagram with $H_{c2}$ and the second
transition including its hysteresis is shown in Fig.\ \ref{figure3}. At
high temperatures, $H_{c2}$ increases very steeply, before it levels
off towards saturation at lower $T$. This clearly signals the crossover
from an orbital $T_c$ reduction at low fields towards a Pauli-%
paramagnetic limitation at higher fields. From the initial
critical-field slope, $H^\prime_{c2} =$ d$H_{c2}/$d$T$, the
orbital-limiting field, $H_{orb}$, can be estimated. Using the $T_c$
reduction of about 0.4~K at 8~T, i.e., $H^\prime_{c2} = 20$~T/K, we
obtain $H_{orb} = 0.7H^\prime_{c2} T_c \approx 130$~T, much larger
than the low-temperature $H_{c2}$ we observe.

The large value of $H_{orb}$ results from the strongly reduced orbital
currents for the field orientation parallel to the layers. For that
reason the field-induced spin polarization becomes important resulting
in the rapid $T_c$ reduction. For $\kappa$-(BEDT-TTF)$_2$Cu(NCS)$_2$,
the Pauli-limiting field can be determined quite accurately from
$H_P = \Delta_0/(\sqrt{2}\mu_B)$ \cite{Clo62}, where $\mu_B$ is the
Bohr magneton and the superconducting energy gap, $\Delta_0$, is well
known from specific-heat studies \cite{Mul02,Wos03}. Using $\Delta_0/
k_BT_c = 2.4$ \cite{Wos03}, we obtain $H_P = 23$~T which agrees very
well with the observed limitation of $H_{c2}$ towards low temperatures
(Fig.\ \ref{figure3}). Indeed, the crossover of the broadened
superconducting transition to a sharp first-order transition shows
that the Pauli-limiting field is reached at $\sim$21~T.

Above this field, the $H_{c2}$ line clearly increases its slope and the
transition develops a latent heat. Simultaneously, the second transition
line appears within the superconducting phase. This strongly suggests
the evolution of the FFLO state in $\kappa$-(BEDT-TTF)$_2$Cu(NCS)$_2$.
Indeed, this superconductor fulfils all requirements necessary for the
formation of the FFLO state. First, it is a strongly type-II superconductor
with a large Ginzburg--Landau parameter $\kappa$ of $100 - 200$ even
for perpendicular fields \cite{Lan04}. Second, the Maki parameter,
$\alpha = \sqrt{2}H_{orb}/H_P \approx 8$ is more than four times larger
than required \cite{Gru66}. Finally, $\xi$ is much smaller than the
mean-free path, $\ell$. The latter was determined for the investigated
sample by means of dHvA measurements. In accordance with well established
literature data \cite{Wos96}, two dHvA frequencies were resolved. The
lower frequency, $F_\alpha = 600(1)$~T, originating from a hole orbit,
together with the measured effective mass, $m_c^\alpha = 3.05(10)m_e$,
allows to calculate the Fermi energy $\epsilon_F^\alpha = 23$~meV, where
$m_e$ is the free-electron mass. For the larger breakdown orbit
[$F_\beta = 3870(20)$~T, $m_c^\beta = 6.5(2)m_e$], comprising all
electrons at the Fermi level, the Fermi energy is $\epsilon_F^\beta =
69$~meV. The field dependence of the dHvA amplitude gives the scattering
time $\tau = 1.9\times 10^{-12}$~s for the $\alpha$ orbit. The magnetic
breakdown allows the determination of the scattering time for the $\beta$
orbit only with much larger error bars. Assuming an unchanged $\tau$
leads to $\ell = 100$ and 115~nm, respectively. By use of $H_{c2\perp} =
4$~T from our specific-heat data [Fig.\ \ref{fig1ab}(a)], $\xi \approx
9$~nm. Correspondingly, $\ell/\xi \approx 12$ proves that our sample is
in the clean limit.

We may now compare our results with the calculated phase diagram
for magnetic fields parallel to the layers. For that, we assume
$s$-wave superconductivity \cite{Mul02,Wos03,Els00}, although
that is not an essential ingredient.
Detailed studies \cite{Shi97,Bur94} showed that the transition lines,
the orders of the transitions, and the structures of the equilibrium
states depend very sensitively on the electronic structure, i.e.,
on the Fermi surface, the effective masses of the quasiparticles,
and their interactions. We model the system under consideration
by a stack of 2D superconducting planes with negligibly small
conductivity perpendicular to the planes and use the Fermi surface
and effective masses derived from the dHvA experiments. As the magnetic
susceptibility in the normal state is not significantly renormalized
compared to the expected value for free quasiparticles we neglect
Landau's spin-dependent interaction parameter $F_{0}^{a}$. For a
quantitative comparison with experiment we have to account for
corrections due to the strong electron-phonon coupling. Considering
the rather small ratio $(k_BT_c)/(\hbar \left\langle \omega \right\rangle
)\simeq 0.08$ \cite{Hag03} we anticipate neither pronounced anomalies in
the variation with $T$ of the enhancement factors nor significant
differences in critical-field renormalizations corresponding to first and
second-order transitions, respectively \cite{Rai74}. As we are mainly
interested in the low-$T$ behavior we implicitly account for
strong-coupling corrections by rescaling the weak-coupling
results with the enhancement factor for the low-$T$ energy
gap $\Delta_0/(1.76k_BT_c)\simeq 1.35$ \cite{Wos03}. The solid line
in Fig.\ \ref{figure3} displays the resulting critical field for the
second-order transition from the normal to the inhomogeneous
FFLO state which agrees well with experiment. The anisotropy of the
effective masses which differ by a factor of $\sim$2 on the different
Fermi-surface sheets stabilizes the inhomogeneous superconducting state
and leads to the steep upturn in $H_{c2}(T)$. Following Ref.\ \cite{Com02}
we find that the transition is of second order in the vicinity of the
tricritical point in close analogy to the case of the isotropic 2D
superconductor with non-interacting quasiparticles. For a better
description of the low-$T$ $H_{c2}(T)$ data and for estimating the
transition from the homogeneous superconducting to the FFLO state
more detailed information on the electronic parameters is necessary.

When we compare the phase diagram in Fig.\ \ref{figure3} with that
of CeCoIn$_5$ \cite{Bia03} we find clear qualitative differences.
For CeCoIn$_5$, $H_{c2}$ is less reduced at high fields due to Pauli
limitation and there is no clear upturn of $H_{c2}$ when the FFLO
state appears. The latter may be due to the rather isotropic
in-plane effective masses. For the 2D organic material, the FFLO phase
occupies only a small area in the phase diagram with the transition
line following closely the $H_{c2}$ transition (Fig.\ \ref{figure3}).
Another difference is found in the nature of the transitions. For the
organic superconductor both transitions are found to be first order,
whereas for CeCoIn$_5$ only the $H_{c2}$ line is first order. Both
materials have similar values for $\alpha$ and $\ell/\xi$ (see \cite{Mic06}
for recent estimates). A possible reason for these differences might be
the anisotropy which for CeCoIn$_5$ is much smaller than for
$\kappa$-(BEDT-TTF)$_2$Cu(NCS)$_2$. For the latter the ratio of the Fermi
energy with respect to the interlayer hopping, $\epsilon_F/t_\perp$, is
about 3700 \cite{Sin02}. Finally, we see no evidence for an FFLO state
in the field perpendicular orientation, whereas for CeCoIn$_5$ the
matter is still unresolved \cite{Bia03,Bia02,Rad03}.

In conclusion, we presented high-resolution specific-heat data in
high magnetic fields that give clear thermodynamic evidence for the
existence of a narrow additional superconducting phase in the 2D
organic superconductor $\kappa$-(BEDT-TTF)$_2$Cu(NCS)$_2$ for
in-plane magnetic fields. This phase is most probably a realization
of the long-time predicted FFLO state. This is supported by the
paramagnetic limitation of $H_{c2}$, the upturn of $H_{c2}$ when
the Pauli-limiting field of 21.5~T is reached, and the first-order
nature of the transitions above this field. The clear observation
of the FFLO state in a second superconductor, besides CeCoIn$_5$,
with a qualitative different phase diagram allows for thorough
tests of our fundamental understanding of superconductivity at
high magnetic fields.

We thank A. Junod, P. Fulde, A.D. Bianchi, T. Giamarchi, \O. Fisher,
and C. Berthier for stimulating discussions and S. Debrion, I. Sheikin,
T.T. Lortz, and the technicians of the Grenoble High Field
Laboratory for their experimental support.

\end{document}